\documentclass[11pt,a4paper]{article}
\usepackage[T1]{fontenc}
\usepackage[utf8]{inputenc}
\usepackage{latexsym, amssymb, enumerate, amsmath, cite, tabu}

    \usepackage{graphicx}
\usepackage{color, cases,empheq}
\usepackage{authblk}
\newcommand{\tildeq}{{\tilde{Q}}}
\newcommand{\Amax}{{A_{\text{max}}}}
\newcommand{\Amin}{{A_{\text{min}}}}
\newcommand{\Abar}{{\bar{A}}}

\newcommand{\Hbar}{{\bar{H}}}

\begin{document}

\title{Nonlinear traveling waves for the skeleton of the Madden--Julian oscillation}
  \author[1]{Shengqian Chen \footnote{Corresponding author}}
  \author[2]{Samuel N. Stechmann} 
  \affil[1]{Department of Mathematics, University of Wisconsin -- Madison, Madison, Wisconsin, USA, sqchen@math.wisc.edu} 
  \affil[2]{Department of Mathematics, and Department of Atmospheric and Oceanic Sciences, University of Wisconsin -- Madison, Madison, Wisconsin, USA, stechmann@wisc.edu}
  \title{Multiscale asymptotics for the Skeleton of the Madden-Julian Oscillation and Tropical--Extratropical Interactions}
  \date{}






\maketitle 
\begin{abstract}
The Madden--Julian Oscillation (MJO) is the dominant component
of intraseasonal (30--90 days) variability in the tropical atmosphere.
Here, traveling wave solutions are presented for the MJO skeleton model
of Majda and Stechmann.
The model is a system of nonlinear partial differential equations
that describe the evolution of the tropical atmosphere on
planetary (10,000--40,000 km) spatial scales.
The nonlinear traveling waves come in four types,
corresponding to the four types of linear wave solutions,
one of which has the properties of the MJO.
In the MJO traveling wave, the convective activity
has a pulse-like shape, with a narrow region of enhanced convection
and a wide region of suppressed convection.  Furthermore,
an amplitude-dependent dispersion relation is derived,
and it shows that the nonlinear MJO has a lower frequency and
slower propagation speed than the linear MJO.  By taking the
small-amplitude limit, an analytic formula is also derived for the
dispersion relation of linear waves.  To derive all of these results,
a key aspect is the model's conservation of energy, which holds
even in the presence of forcing.  In the limit of weak forcing,
it is shown that the nonlinear traveling waves have a simple
sech-squared waveform.

\end{abstract}

%

\section{Introduction}\label{intro}

This paper concerns the following system of nonlinear 
partial differential equations (PDEs):
 \begin{subequations}
 \begin{alignat}{4}
 K_t + K_x &= -\frac{1}{2}(\Hbar A -F) \label{K_t}\\
 R_t -\frac{1}{3}R_x &= -\frac{1}{3} (\Hbar A -F)\label{R_t}\\
 Q_t + \tildeq K_x -\frac{\tildeq}{3}R_x&= (\frac{\tildeq}{6}-1)(\Hbar A -F)\label{Q_t}\\
 A_t &= \Gamma Q A \label{A_t}
 \end{alignat}
 \label{krqa}
 \end{subequations}
which was originally designed in
\cite{ms09pnas}.
This is a hyperbolic system whose only nonlinearity is in equation (\ref{A_t}).
The main goals of this paper are (i) to present nonlinear traveling wave
solutions of (\ref{krqa}) and (ii) to describe the 
features of the nonlinear waves that are absent from the linear waves.
An important element will be conservation of energy, which
holds even in the presence of the source term, $F$.

In equations~(\ref{krqa}), the variables $K$, $R$, $Q$, and $A$
represent the state of the atmosphere near the equator.
$K$ and $R$ represent Kelvin and equatorial Rossby wave circulation patterns,
and they are related to the velocity and temperature as described
further below.
$Q$ represents the lower-tropospheric water vapor 
(``moisture'' hereafter),
and $A$ represents the amplitude of deep convective activity.
As such, $A$ accounts for an important moisture sink and heat source
for the atmosphere, from rainfall and latent heating,
as represented by the proportionality constant
$\Hbar$ for heating.

The system (\ref{krqa}) was proposed in \cite{ms09pnas} 
as a model for the Madden--Julian Oscillation (MJO).
The MJO is the dominant component of 
intraseasonal ($\approx$ 30-60 days) variability in the tropical atmosphere.  
This variability appears not only in the wind, pressure, and temperature 
fields, but also in water vapor and precipitation/convection.  
The structure of 
the MJO is a planetary-scale ($\approx$ 10,000-40,000 km) circulation cell 
with regions of enhanced and suppressed convection, and it propagates 
slowly eastward at a speed of roughly 5 m/s. The main regions of 
MJO convective activity are over the Indian and western Pacific Oceans, and the 
MJO interacts with monsoons, tropical cyclones, 
El Ni\~{n}o--Southern Oscillation, and other 
tropical phenomena.  Nevertheless, while the MJO is mainly a tropical 
phenomenon, it also interacts with the extratropics and can affect 
midlatitude predictability.    
See 
\cite{mj71,mj72,mj94,z05,lw11}
for further background information on the MJO.

Despite the wealth of studies dedicated to the MJO,
theory and numerical simulations are still major challenges.
Several studies have documented the inadequacies and progress
of general circulation model (GCM) simulations
\cite{setal96,letal06,kimetal09},
and new techniques continue to be developed and show
increasingly realistic results
\cite{g03,br09,ksmt11,akm13}.
A large part of the challenge is the complex multiscale structure
of the MJO
\cite{n88,hl94,dltk13}.  
More theoretical work is needed to better understand the
multiscale processes at work in the tropics.
For recent reviews, see
\cite{m10,k10,kms13}.

Motivated by the MJO's multiscale structure,
the terminology of the MJO's ``skeleton'' and ``muscle''
was introduced in \cite{ms09pnas}, and it
refers to the MJO's intraseasonal--planetary-scale envelope 
and further details beyond the envelope, respectively.
Further work with the MJO skeleton model can be found in
\cite{ms09pnas,ms11,tms14},
including a stochastic version of the model \cite{tms14}.
Work on the MJO's ``muscle'' can be found in
\cite{mb04,bm05,ms09cmtJAS,khms12,sms13},
which focus on the role of convective momentum transport
\cite{mk97}.

In addition to (\ref{krqa}),
which includes effects of the Coriolis force
and meridional ($y$) variations of the circulation,
a simpler yet less realistic system will also be considered here:
\begin{subequations}
\begin{alignat}{4}
u_t - \theta_x &=0 \label{u_t}\\
\theta_t-u_x &= \Hbar a-F \label{th_t}\\
q_t + \tildeq u_x &= -\Hbar a + F \label{q_t}\\
a_t &= \Gamma q a \label{a_t}
\end{alignat}
\label{eq1d}
\end{subequations}
where $u$ is a velocity and $\theta$ represents a temperature
\cite{ms09pnas}.
This system neglects meridional ($y$) variations 
and can be thought of as the atmospheric circulation
directly above the equator, $y=0$, where the
Coriolis force vanishes.
Many of the results here will hold equally well for
(\ref{krqa})
and
(\ref{eq1d}).
While 
(\ref{eq1d})
neglects important physics,
their east--west symmetry is mathematically advantageous,
as it provides a simplification of the equations.
Also, these equations have the advantage of
being written in terms of zonal velocity $u$ and potential
temperature $\theta$,
which are more physically intuitive than the wave amplitude variables $K$ and $R$.

The nonlinear traveling waves of the present paper
are an addition to several solitary wave systems
for other atmospheric phenomena.
Several examples take the form of coupled KdV systems,
including some with nonlinear self-interaction and linear coupling
\cite{m94,gg99}
and some with nonlinear coupling
\cite{mb03,b09camb,bm04sam,o09,gst13}.
The linearly coupled systems \cite{m94,gg99}
were derived in the context of midlatitude baroclinic instability,
and the nonlinearly coupled system \cite{mb03}
was derived in the context of tropical--extratropical interactions
of equatorial and midlatitude Rossby waves.
What physically distinguishes the present MJO nonlinear waves
from these coupled KdV systems is  that, the MJO skeleton model has coupling
with moisture and convection,
whereas for the KdV-like systems for barotropic and baroclinic instabilities,
the nonlinearities come from the transport terms of momentum and temperature \cite{mb03}.

In a model with a different treatment of convection,
precipitation front solutions have been investigated
as traveling waves with a discontinuous transition
\cite{fmp04}
or a steep gradient 
\cite{sm06}
between a precipitating region and a non-precipitating region.
Unlike (\ref{krqa}),
the precipitation front equations include a nonlinear switch
(Heaviside function) in the precipitation term,
and the system has the mathematical form 
of a hyperbolic free boundary problem
\cite{ms10}.

In the MJO skeleton model (\ref{krqa}),
the only nonlinear interaction is in the coupling of
moisture $Q$ and convective activity $A$.
As described in more detail below,
the moisture--convection coupling has a mathematical form
that is reminiscent of the nonlinearity in the 
Toda lattice model
\cite{t67,t75}
when written in terms of Flaschka's variables
\cite{f74}.

The rest of the paper is organized as follows. 
Section~\ref{dsc-sec} describes the physical mechanism and simplifying assumptions for the model. 
The model conserves a total energy even in the presence of the forcing term, 
which is crucial to have an analytical waveform.
In section~\ref{nonlinear-sec}, the nonlinear traveling wave solutions are presented. 
The properties of the nonlinear waves are compared with those of their linear analogues,
 including the traveling wave speed,
the dispersion relation, and the shape of solutions. 
In section~\ref{phys-sec}, physical quantities are recovered from variables $K$, $R$, $Q$ and $A$, 
and their physical significances in tropical climate are stated.
Section~\ref{future-sec} provides some further explorations of the model:
 the stability/instability of traveling wave solutions, 
the sech-squared waveform under the weak forcing limit, 
and the key results for the east-west symmetric system~(\ref{eq1d}).

\section{Model description and energetics}\label{dsc-sec}
In this section, physical mechanisms and assumptions are described for the MJO skeleton model.
\subsection{Model description}\label{dsc}

The MJO skeleton model was originally proposed and developed in \cite{ms09pnas}. 
It is a nonlinear oscillator model for the MJO skeleton as a neutrally stable wave,
i.e., the model includes neither damping nor instability mechanisms.

To obtain the simplest model for the MJO, truncated vertical and meridional structures are used.
For the vertical truncation, only the first baroclinic mode is used 
so that $u(x,y,z,t)=\sqrt{2}u^*(x,y,t)\cos(z)$, etc \cite{m03, bm06dao}.
The stars are dropped to keep the expression simple:
\begin{subequations}
\begin{alignat}{2}
u_t -y v - \theta _x &=0 \label{2d-ut}\\
yu -\theta_y &= 0 \label{2d-vt}\\
\theta_t -u_x-v_y &=\Hbar a - F\label{2d-tt} \\
q_t - \tildeq (u_x+v_y) &= -\Hbar a +F \label{2d-qt}\\
a_t &= \Gamma q a. \label{2d-at}
\end{alignat}
\label{2d-eq}
\end{subequations}
The model (\ref{2d-eq}) is a nondimensional model, with the scaling listed in table~\ref{tb_nondim}, 
taken from \cite{sm06}.
In this paper, the nondimensional variables are used throughout the derivations and calculations.
Here, the coordinate system $(x,y,z)$ represents zonal, meridional and vertical directions.
For the meridional coordinate, typically, $y=0$ is located at the equator, where the latitude is~0.
For the vertical coordinate, $z=0$ and ${\pi}$ are located at the bottom and top of the troposphere.

\begin{table}
  \centering 
  \begin{tabu}{llll}
\tabucline[1pt]{-}
Par. & Derivation & Dim. val. & Description\\
\hline
$\beta$ & & $2.3\times10^{-11}$~m$^{-1}$s$^{-1}$ & Variation of Coriolis parameter with latitude\\
$\theta_0$ && $300$~K & Potential temperature at surface \\
$g$ && $9.8$~m~s$^{-2}$ & Gravitational acceleration\\
$H$ && $16$~km & Tropopause height\\
$N^2$ & $(g/\theta_0)\mathrm{d}\bar{\theta}/\mathrm{d}z$ & $10^{-4}$~s$^{-2}$ &
Buoyancy frequency squared\\
$c$ & $NH/\pi$ & $50$~m~s$^{-1}$ & Velocity scale\\
$X_e$ & $\sqrt{c/\beta}$ & $1500$~km & Equatorial length scale\\
$T$ & $X_e/c$ &$8$~hrs & Equatorial time scale\\
       & $HN^2\theta_0/(\pi g)$ & $15$~K & Potential temperature scale \\
       & $H/\pi$ & $5$~km & Vertical length scale\\
       & $H/(\pi T)$ & $0.2$~m~s$^{-1}$ & Vertical velocity scale \\
       & $c^2$ & $2500$~m$^2$~s$^{-2}$ & Pressure scale\\
    \tabucline[1pt]{-}
\end{tabu}
  \caption{ From \cite{sm06}. Constants and reference scales for nondimensionalization. }
  \label{tb_nondim}
\end{table}

In (\ref{2d-eq}), $u$ and $v$ are zonal and meridional velocities, respectively;
and $\theta$ is the potential temperature.
In the 2D shallow water system (\ref{2d-eq}), the dry dynamical core of the model (\ref{2d-ut})-(\ref{2d-tt}) is the equatorial long-wave equations \cite{g82, mb04, bm05, bm06dao, m03}. 
The long-wave assumption is based on the fact that planetary equatorial waves have long zonal wavelength ($\sim$15,000~km),
comparing to their spans in the meridional and vertical directions ($\sim$1,500~km).
In the zonal long-wave limit, the $v_t$ term is neglected\cite{mk03,m03}
and high frequency inertia-gravity waves are filtered out. 
Another remark is that the $\beta$-plane approximation is applied for Coriolis force at the equator,
where $\sin(y)\sim y$ as $y\to 0$.

While $u$, $v$ and $\theta$ are from the dry dynamics, the other 2 variables are included to represent moist convective processes: 
\begin{equation}
\begin{aligned}
&\mbox{$q$: lower tropospheric moisture }\\
  &\mbox {$a$: amplitude of wave activity envelope}
  \end{aligned}
\end{equation}
The nondimensional dynamical variable $a$ parameterizes the amplitude of the planetary scale 
envelope of synoptic scale wave activity.

A key part of the $q-$and$-a$ interaction is how the moisture anomalies influence the convection.
The premise is that, for convective activity on planetary/intraseasonal scales,
 it is the time tendency of convective activity, not the convective activity itself,
 that is most directly related to the lower-tropospheric moisture anomaly.
 In other words, rather than a functional relationship $a=a(q)$,
  it is posited that $q$ mainly influences the tendency, 
  i.e., the growth and decay rates, of the convective activity.
  The simplest equation that embodies this idea is (\ref{2d-at}),
where $\Gamma$ is a constant of proportionality:
positive (negative) low-level moisture anomalies create a tendency to enhance (decrease)
the envelope of convective wave activity.
The basis for (\ref{2d-at}) is supported by a combination of observations, modeling, and theory 
(see \cite{ms09pnas} and references therein for more information).


%


Notice this model contains a minimal number of parameters: $\tildeq = 0.9$, 
the (nondimensional) mean background vertical moisture gradient;
$\Gamma=1$, 
or $\approx 0.5$~d$^{-1}$(g$\cdot$kg$^{-1}$)$^{-1}$ dimensionally, 
where $\Gamma q$ acts as a dynamic growth/decay rate of the wave activity envelope;
$F=0.023$, or $\approx1$~K$/$d in dimensional unit, 
is the fixed, constant radiative cooling rate;
and $\Hbar = 0.23$, or $\approx10$~K$/$d in dimensional unit,
is a constant heating rate prefactor.
Note that $F$ is taken to be a constant here for simplicity, but
it could also be chosen to be a function of $x$ and/or $t$.
Also note that $\Hbar$ can be scaled out of equation by rescaling the ``a''
variable. However, to maintain consistency with model presentations
in literature \cite{ms09pnas,ms11}, we find it favorable to write the model in the same fashion.

Next, the model (\ref{2d-eq}) is projected and truncated at leading parabolic cylinder functions $\Phi_m(y)$\cite{m03, bm06dao}. 
The parabolic cylinder functions $\left\{\Phi_m(y)\right\}_{m=0}^{\infty}$ form an orthonormal basis in the meridional ($y$) direction, with respect to $L^2$, where the inner product is defined by
\begin{equation}
\langle f, g\rangle = \int_{-\infty}^{+\infty} f(y) g(y) \mathrm{d}y.
 \end{equation}
In system (\ref{2d-eq}), the variables can be written as projections to the parabolic cylinder functions.
For example,
\begin{equation}
u(x,y,t) = \sum _{m=0}^{\infty} u_m(x,t) \Phi_m(y),
\end{equation} 
where
\begin{equation}
u_m(x,t) = \langle u(x,y,t), \Phi_m(y)\rangle.
\end{equation}
 The leading parabolic cylinder functions are
 \begin{equation}
\begin{aligned}
\Phi_0(y)&= \pi^{-1/4}\exp(-y^2/2),  \\
\Phi_1(y)&= \pi^{-1/4}\sqrt{2}y\exp(-y^2/2),  \\
\Phi_2(y)&= \pi^{-1/4}2^{-1/2}(2y^2-1)\exp(-y^2/2).
\end{aligned}
\end{equation}
In the derivation of (\ref{krqa}), it is assumed that $a$, 
 the envelope of convective wave activity, has a simple equatorial meridional structure
 proportional to $\Phi_0(y)$:
 $a(x,y,t)=A(x,t)\Phi_0(y)$.
 Such a meridional heating structure excites only Kelvin waves 
 and the first symmetric equatorial Rossby waves \cite{bm06dao, m03, tms14},
 and the resulting meridionally truncated equations are written in (\ref{krqa}).
 The nondimensional heating rate prefactor $\Hbar$, is set to be $\Hbar = 0.23$. 
 
 The meridional projections of the velocity and potential temperature fields
 take the form
  \cite{bm06dao, m03, ms09pnas,ms11,tms14}
\begin{equation}
\begin{aligned}
 u(x,y) &= \left[K(x)-R(x)\right]\Phi_0(y)+\frac{\sqrt{2}}{2}R(x)\Phi_2(y),\\
v(x,y)&=  \frac{1}{3\sqrt{2}}\left[4R'(x)-(\Hbar A(x)-F)\right]\Phi_1(y), \\
\theta(x,y)&=  -\left[K(x)+R(x)\right]\Phi_0(y)-\frac{\sqrt{2}}{2}R(x)\Phi_2(y).
\end{aligned}
\label{reform}
\end{equation}
where $R$ and $K$ are the contributions from the Rossby and Kelvin waves, respectively.
The meridional structure of $q$ is given by $q(x,y,t)=Q(x,t)\Phi_0(y)$.
Note that the definitions of $K$ and $R$ are different than in \cite{ms09pnas}; here,
$K$ and $R$ have been scaled by $\frac{1}{\sqrt{2}}$ and $\frac{1}{2\sqrt{2}}$,
respectively, as was also done in \cite{tms14}. 

\subsection{Energetics}
The nonlinear MJO skeleton model has an important energy principle: 
the system (\ref{2d-eq}) conserves a total energy that includes a contribution from the convective activity $a$:
\begin{equation}
\partial_t\left[ \frac{1}{2}u^2 + \frac{1}{2}\theta^2 +  \frac{\tildeq}{2(1-\tildeq)}\left(\theta+ \frac{q}{\tildeq}\right)^2 +  \frac{\Hbar}{\Gamma \tildeq} a - \frac{F}{\Gamma \tildeq} \log a \right]
-  \partial_x\left(u\theta \right) -\partial_y\left(v\theta \right)=0
\label{consE_2d}
\end{equation}
This total energy is a sum of contributions from 
dry kinetic energy $\frac{1}{2}u^2$, potential energy $ \frac{1}{2}\theta^2$, 
moist potential energy $\frac{\tildeq}{2(1-\tildeq)}\left(\theta+ \frac{q}{\tildeq}\right)^2$, and 
convective energy $ \frac{\Hbar}{\Gamma \tildeq} a - \frac{F}{\Gamma \tildeq} \log a$. 
This energy conservation also holds for (\ref{eq1d}), 
where the meridional ($y$) variation is neglected. 
In this case, the term $\partial_y\left(v\theta \right)$ disappears.

Likewise, the system (\ref{krqa}) conserves a total energy :
\begin{equation}
\partial_t\left[K^2 + \frac{3}{2}R^2 +\frac{\tildeq}{2(1-\tildeq)}\left(\frac{Q}{\tildeq}-K-R\right)^2+ \frac{\Hbar}{\Gamma \tildeq} A - \frac{F}{\Gamma \tildeq} \log A\right] + \partial_x\left(K^2-\frac{1}{2}R^2\right)=0.
\label{consE_total}
\end{equation}	
This total energy is a sum of contributions from 
dry kinetic and potential energy $K^2 + \frac{3}{2}R^2 $, 
moist potential energy $\frac{\tildeq}{2(1-\tildeq)}\left(\frac{Q}{\tildeq}-K-R\right)^2$, and 
convective energy $\frac{\Hbar}{\Gamma \tildeq} A - \frac{F}{\Gamma \tildeq} \log A$. 
Note that the natural requirement on the background moisture gradient,
$0<\tildeq<1$, is needed to guarantee a positive moist potential energy.
The convective energy part $\frac{\Hbar}{\Gamma \tildeq} A - \frac{F}{\Gamma \tildeq} \log A$ achieves its minimum value at the radiative-convective equilibrium state, i.e. $A=\Abar$, where
\begin{equation}
\Abar={F}/{\Hbar}=0.1.
\label{abar_def}
\end{equation}

While (\ref{consE_2d}) is the total energy conservation for system (\ref{2d-eq}),
another conserved quantity is readily obtained by summing up (\ref{2d-tt})-(\ref{2d-qt})
to eliminate source and forcing terms.
The conservation law is given by
\begin{equation}
\partial_t(\theta+q) - (1-\tildeq)(\partial_x u + \partial_y v)=0
\end{equation}
The quantity $\theta+q$ is an analogue of the moist static energy.

\section{Nonlinear traveling wave solutions}\label{nonlinear-sec}
In this section, a traveling wave ansatz is applied to system (\ref{krqa}) 
and exact solutions can be found, with certain restrictions on wave speed.
Based on the exact traveling wave formula, connections are built to relate wave amplitude, wavelength,
traveling speed and total energy.
The nonlinear solutions are compared with linear solutions, 
demonstrating amplitude-dependent features.

\subsection{Reduction to nonlinear oscillatory ODE}
With the assumption that the wave travels with speed $s$,
the traveling wave ansatz converts the PDE system (\ref{krqa}) to a set of ODEs,
by writing $[K,R,Q,A](x,t) = [K,R,Q,A](\tilde{x})$
where $\tilde{x}=x-st$:
\begin{subequations}
\begin{alignat}{4}
(-s+1)K' &= -\frac{1}{2}(\Hbar A-F) \label{dK}\\ 
(-s-\frac{1}{3})R' &= -\frac{1}{3}(\Hbar A -F) \label{dR} \\
-sQ' +\tildeq K' -\frac{\tildeq}{3}R' &=\left(\frac{\tildeq}{6}-1\right)(\Hbar A-F) \label{dQ}\\
-sA' &= \Gamma Q A. \label{dA}
 \end{alignat}
 \label{dkrqa}
 \end{subequations}
From (\ref{dK}) and (\ref{dR}), $K'$ and $R'$ in (\ref{dQ}) can be replaced by 
\begin{equation}
K' = \frac{1}{2(s-1)}(\Hbar A-F)  \quad \text{and} \quad R' = \frac{1}{3s+1}(\Hbar A-F),
\label{KR-replace}
\end{equation}
which further simplifies the ODE system to 
 \begin{subequations}
 \begin{alignat}{2}
 Q' &= \frac{f(s)}{6s}(\Hbar A-F)\label{simpleQ}\\
 A' &= -\frac{\Gamma}{s}Q A \label{simpleA}
 \end{alignat}
 \label{simpleAQ}
 \end{subequations}
 where
 \begin{equation}
  f(s)=\frac{3\tildeq}{s-1}-\frac{2\tildeq}{1+3s}-\tildeq+6.
  \label{fs-eq}
  \end{equation}
  The plot of $f(s)$ is shown in figure~\ref{fs-fig}.
  The nonlinearity in (\ref{simpleAQ}) is reminiscent of the Toda lattice model \cite{t67,t75}
  when written in terms of Flaschka's variables \cite{f74}.
  Using this connection, the change of variables 
   $B=\log A$
   transforms (\ref{simpleAQ}) into 
     a Hamiltonian system that 
    has been called the Toda oscillator \cite{op85, ldv02, kblu07}.
  The Hamiltonian function for the Toda oscillator 
  is a conserved quantity for (\ref{simpleAQ}):
\begin{equation}
 \mathcal{H}(Q,A)=\frac{1}{s}\left[\frac{\Gamma}{2}Q^2 + \frac{f(s)}{6}(\Hbar A-F \log A ) \right].
 \label{conserveE}
\end{equation}
The function $\mathcal{H}(Q,A)$
 is plotted in figure~\ref{Es-fig},
 and it will play an important role in the derivations that follow.
 In particular,
 periodic orbits of the Hamiltonian ODEs correspond to periodic traveling waves of (\ref{krqa}).
\subsection{Allowed traveling wave speed}\label{exist-sec}
\begin{figure}
\centering
 \includegraphics[width=.32\textwidth]{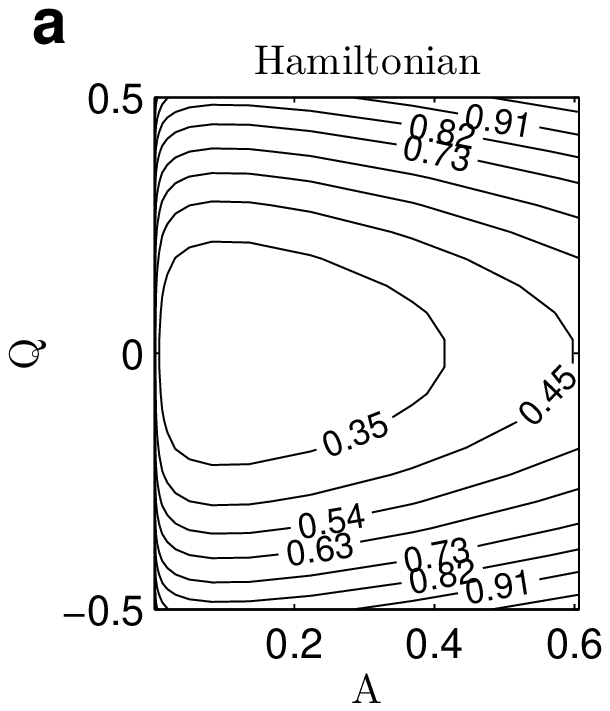}
\includegraphics[width=.32\textwidth]{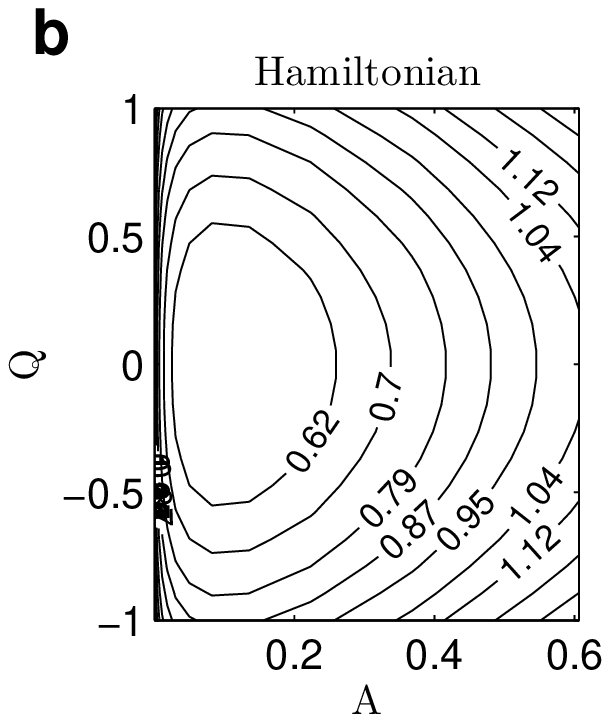}
\includegraphics[width=.32\textwidth]{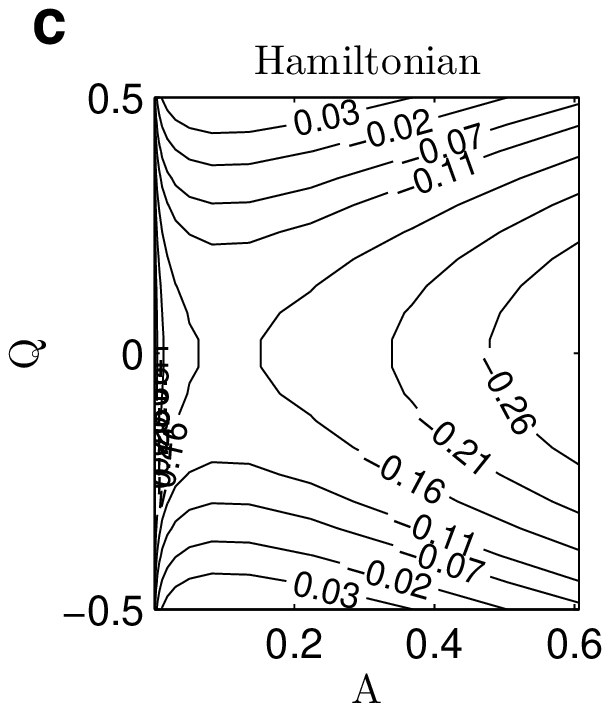}
\caption{Contour plots of Hamiltonian function $\mathcal{H}$ from (\ref{conserveE}) with different choices of traveling wave speed $s$. (a): $s=0.2$, $f(s)>0$. The critical point $[Q_0,A_0]$ is a minimum for $\mathcal{H}$; (b): $s=1.5$, $f(s)>0$. The critical point $[Q_0,A_0]$ is a minimum for $\mathcal{H}$; (c): $s=0.5$, $f(s)<0$. the critical point $[Q_0,A_0]$ is a saddle for $\mathcal{H}$.}
\label{Es-fig}
 \end{figure}

The periodic traveling wave solutions exist only for certain wave speeds. 
Solutions are point values $[Q,A]$ on the closed contours of Hamiltonian function $\mathcal{H}$
from (\ref{conserveE}).
To form closed contours, the function $\mathcal{H}$ needs to be convex in both $Q$ and $A$, 
which is equivalent to the positivity of $f(s)$:
\begin{equation}
f(s)>0.
\label{fs-crit}
\end{equation}
Under the condition (\ref{fs-crit}), the critical point of system (\ref{simpleAQ}), 
\begin{equation}
[Q_0,A_0]=[0,\Abar],
\end{equation}
is a local extreme for $\mathcal{H}$ (figure~\ref{Es-fig}a,b),
and there are closed contours. 
On the other hand, with $f(s)<0$, 
the critical point is a saddle (figure~\ref{Es-fig}c). 

\begin{figure}
 \centering
 \includegraphics[width=.6\textwidth]{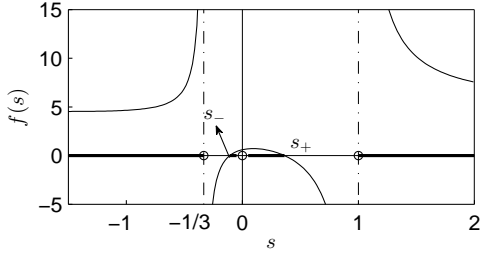}
 \caption{Plot of $f(s)$ from (\ref{fs-eq}) with $\tildeq = 0.9$. The thick lines are four groups of eligible traveling wave speed $s$ that allow for traveling wave solutions. From left to right, they correspond to four modes:
 dry Rossby, moist Rossby, MJO and dry Kelvin.}
 \label{fs-fig}
 \end{figure}

The positivity requirement (\ref{fs-crit}) is met by four groups of traveling wave speed $s$
as shown in figure~\ref{fs-fig}:
\begin{equation}
\begin{aligned}
s<-\frac{1}{3}: \quad & \text{dry Rossby,}\\
 s_-<s<0: \quad & \text{moist Rossby,}\\
  0<s<s_+: \quad & \text{MJO,}\\
   s>1: \quad & \text{dry Kelvin,}
\end{aligned}
\label{s-crit}
\end{equation}
where $s_{\pm}$ are roots for $f(s)=0$. 
The four groups of traveling wave speeds are analogous to four eigenvalues in the linearized system~\cite{ms09pnas}.

Note that although the positivity of $f(s)$ gives a wide range of eligible traveling wave speeds,
realistically, when the equatorial circumference is considered, the traveling wave speeds will 
be confined by the longest wavelength that are allowed, 
which will be further discussed in Section~\ref{phys-sec}.

 In (\ref{s-crit}), the boundary values, $s_{\pm}$, set limits for traveling wave speeds of two moist modes.
  They depend on the moisture gradient coefficient $\tildeq$.
 The dry modes, on the other hand, are not confined by $\tildeq$.
 Similar results can be found in the precipitation front models, e.g.\cite{fmp04, sm06}.
 
\subsection{Analytical waveform}\label{sol-sec}
With the choice of wave speed $s$ satisfing (\ref{s-crit}), 
analytical waveforms are obtained.
A closed contour of $\mathcal{H}$ determines the particular waveform,
and the contour is selected by any given extreme values of either $A$ or $Q$.
According to (\ref{conserveE}), for any closed contour of $\mathcal{H}$ as in figure~\ref{Es-fig},
when $Q=0$, $A$ achieves its maximum/minimum values, $\Amax$ and $\Amin$.
When $A=\Abar$, $Q$ achieves its extreme values.
While the value for convective wave envelope $A$ is always positive, 
the values of $Q$ shows a positive-negative symmetry.

When the maximum value of $A$ is selected, the Hamiltonian $\mathcal{H}$ is given by (\ref{conserveE}):
\begin{equation}
 \mathcal{H}=\frac{f(s)}{6 \Gamma s}(\Hbar \Amax-F \log \Amax ),
\end{equation} 
For other points $[Q,A]$ on the same contour, they satisfy
\begin{equation}
\frac{1}{s}\left[\frac{1}{2}Q^2 + \frac{f(s)}{6 \Gamma}(\Hbar A-F \log A ) \right] =\frac{f(s)}{6 \Gamma s}(\Hbar \Amax-F \log \Amax ).
\label{prior-ode}
\end{equation}
According to (\ref{simpleA}), $Q$ can be replaced by 
\begin{equation}
Q = -\frac{s A'}{\Gamma A}
\end{equation}
so that (\ref{prior-ode}) becomes an ODE for $A$:
\begin{equation}
\frac{s^2 A'^2}{2 \Gamma^2 A^2}  + \frac{f(s)}{6 \Gamma}(\Hbar A-F \log A ) =\frac{f(s)}{6 \Gamma }(\Hbar \Amax-F \log \Amax ),
\end{equation}
which can be further written as
\begin{equation}
\left(\frac{dA}{d{\tilde{x}}}\right)^2 = \frac{\Gamma f(s)}{3 s^2} A^2 \left[ \Hbar (\Amax-A)-F (\log \Amax - \log A )\right].
\label{ode-now}
\end{equation}
This separable ODE has solution in the implicit form:
 \begin{equation}
  \tilde{x} = \pm  \frac{\sqrt{3 s^2} }{ \sqrt{\Gamma f(s)}}\int_{A}^{A_{\text{max}}}\hat{a}^{-1} \left[ \Hbar (\Amax-\hat{a})-F (\log \Amax -\log \hat{a} )\right]^{-1/2} \mathrm{d}\hat{a} + x_0
  \label{imp_sol}
 \end{equation}
where $\tilde{x}=x-st$ was defined above in (\ref{dkrqa}). 
The integration constant $x_0$ is chosen so that $\Amax$ lies at the center of the domain.
Next, from (\ref{conserveE}), $Q$ is derived:
\begin{equation}
Q(\tilde{x}) = \pm \sqrt{\frac{f(s)}{3\Gamma}}\left\{ \Hbar [\Amax-A(\tilde{x})]-F [\log \Amax -\log A(\tilde{x}) ]\right\}^{\frac{1}{2}},
\label{q_sol}
\end{equation}
where the sign needs to be consistent with (\ref{simpleA}), depending on the growth/decay of $A$ and
the direction of wave propagation.
The other two variables $K$ and $R$ can be obtained by combining (\ref{KR-replace}) and (\ref{simpleQ}):
\begin{equation}
K= -\frac{3s}{(1-s)f(s)} Q,\quad
R = \frac{6s}{(1+3s) f(s)}Q .
\label{kr_sol}
\end{equation}
The integration constants are chosen to be zero so that $K$ and $R$ have zero mean values.
Equations (\ref{imp_sol})-(\ref{kr_sol}) are the analytical traveling waveform for the nonlinear system~(\ref{krqa}).

\begin{figure}
\centering
\includegraphics[width=.8\textwidth]{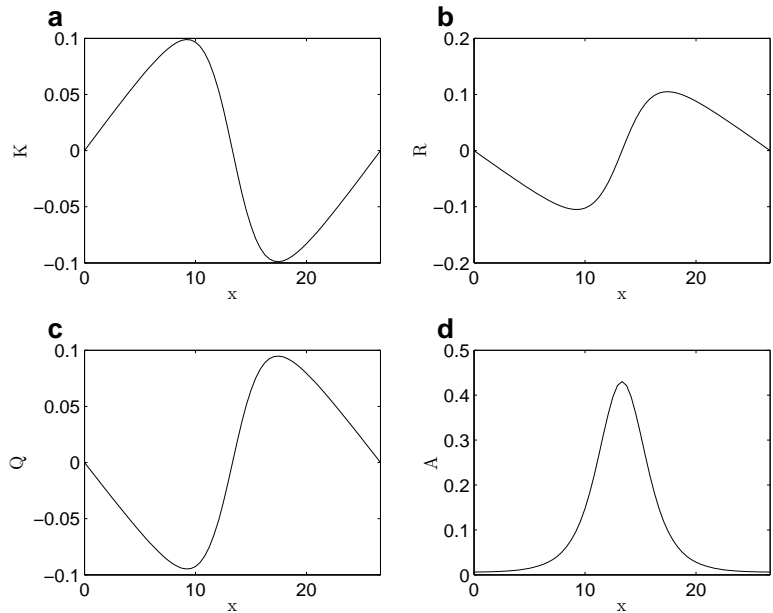}
\caption{Nondimensional MJO mode traveling wave solution in $K$, $R$, $Q$ and $A$. Total energy $\mathcal{E}=3.8$ and wavenumber $k=1$.}
\label{E3_8MJO_fig}
\end{figure}

As an initial illustration, an example of an MJO waveform is shown in figure~\ref{E3_8MJO_fig}.
The main nonlinear feature is that convective activity $A$ has a pulse-like shape:
the region of enhanced convection is narrower than the region of suppressed convection.
At the same time,
the positive anomaly is stronger than the negative anomaly
(recall from (\ref{abar_def}) that the equilibrium value $\bar{A}$ is $0.1$).
What remains to be described is how to construct a waveform
with certain features specified 
-- e.g., with a wavelength $X$ that is a divisor of the Earth's circumference.

\section{Relating wavelength, speed, amplitude and energy}
By using the analytical waveform (\ref{imp_sol}), connections are now built to link
wavelength, speed, amplitude and energy of traveling waves.
Any two of these quantities can determine the other two.
\subsection{Relations between amplitude and maximum/minimum values}
The maximum/minimum values of convection envelope, $\Amax$ and $\Amin$,
are achieved when $Q=0$ in (\ref{conserveE}), so that the following equality holds:
\begin{equation}
\Hbar \Amax - F\log \Amax = \Hbar \Amin - F\log \Amin.
\label{eq_amin_amax}
\end{equation}
The values for $\Amax$ and $\Amin$ can be written in terms of the wave amplitude $\mathcal{A} =\Amax-\Amin$,
so that (\ref{eq_amin_amax}) can be rewritten as
 \begin{equation}
 \Amax = \left[1+(e ^{ \mathcal{A} /\Abar }-1)^{-1}\right] \mathcal{A},
 \quad
 \Amin = (e ^{ \mathcal{A} /\Abar } -1 )^{-1} \mathcal{A},
 \label{A-Amax}
  \end{equation}
where the amplitude $\mathcal{A}$ is defined as:  
$$\mathcal{A}=\Amax-\Amin.$$ 

  \subsection{Amplitude-dependent dispersion relation}
With waveform (\ref{imp_sol}), the wavelength $X$ of the solution can be written as a function of $s$ and $\mathcal{A}$:
\begin{equation}
X= 2\sqrt{ \frac{3s^2}{\Gamma f(s)} }I(\mathcal{A}), 
\label{period_exp}
\end{equation}
 where
 \begin{equation}
 I(\mathcal{A}) = \int_{\Amin}^{\Amax} \hat{a}^{-1} \left[ ( \Hbar(\Amax- \hat{a})-F (\log \Amax - \log \hat{a} )\right]^{-1/2} \mathrm{d}\hat{a} .
 \end{equation}
The function $I(\mathcal{A})$ illustrates that the wavelength 
depends on the amplitude $\mathcal{A}$, in addition to the dependency on wave speeds.
\begin{figure}
\centering
\includegraphics[width=.44\textwidth]{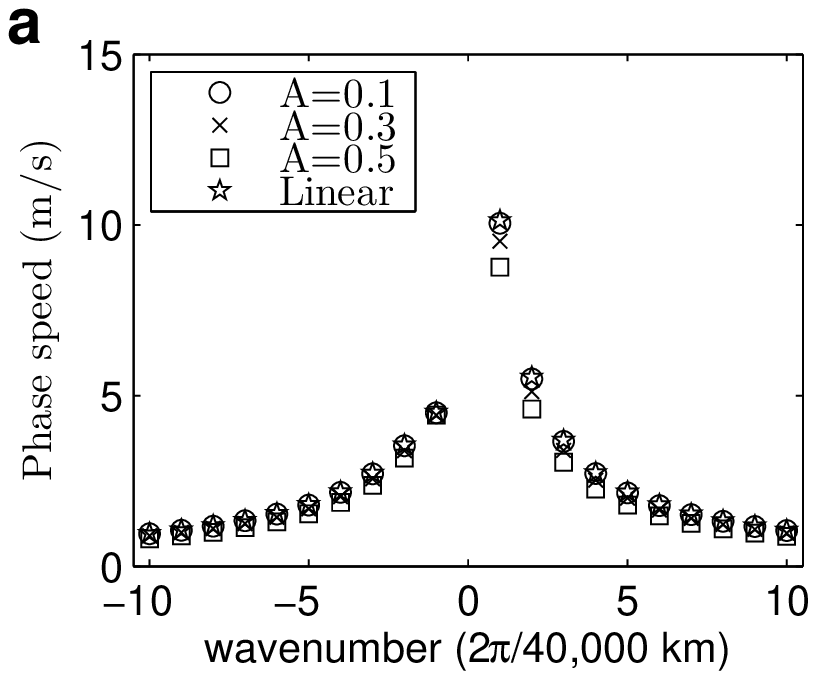}
\includegraphics[width=.45\textwidth]{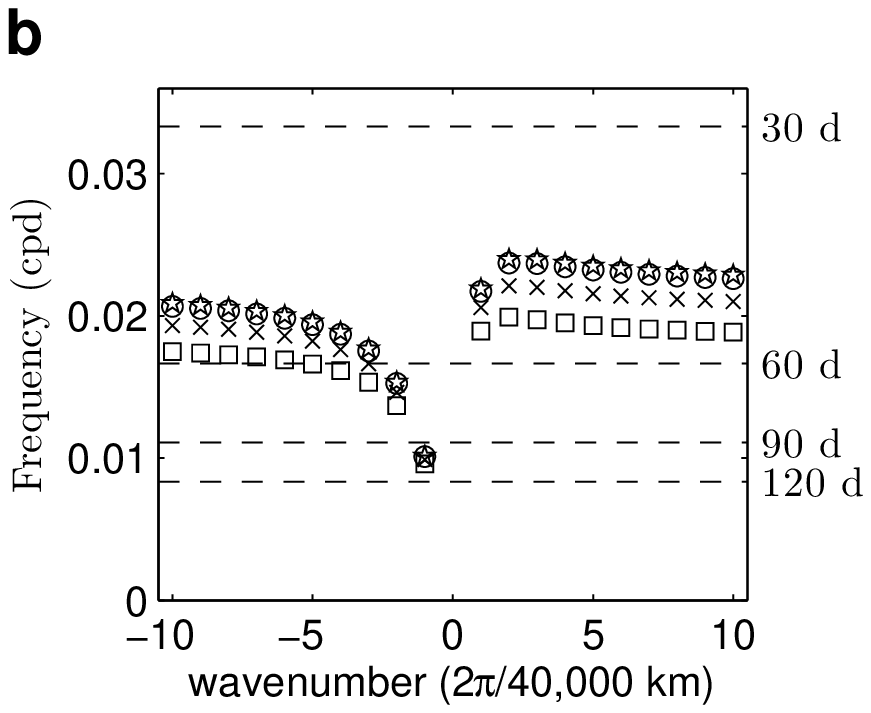}
\caption{Phase speed $s=\omega/k$ (a) and oscillation frequency $\omega(k)$ (b) for nonlinear and linear solutions.
Circle: $\mathcal{A}=0.1$; cross: $\mathcal{A}=0.3$; square: $\mathcal{A}=0.5$; star: linear system.
The large amplitude phase speeds, i.e. $\mathcal{A}=0.5$, are $8.76$~m/s and $4.61$~m/s for $k=1$ and $2$ equatorial waves.
The linear phase speeds, correspondingly, are $10.13$~m/s and $5.55$~m/s.  }
\label{disp_amp_fig}
\end{figure}

In addition to the amplitude-dependent function (\ref{period_exp}) for the nonlinear waves, 
a similar function can be derived in the small-amplitude limit for linear waves.
The amplitude dependency on dispersion relation is a nonlinear feature.
If the system (\ref{krqa}) is linearized around $A=\Abar$, and the traveling wave ansatz is applied,
the simplified ODE system is the linearized (\ref{simpleAQ}) around $A=\Abar$,
 \begin{equation}
 \begin{aligned}
 Q' &= \frac{f(s)}{6s}\Hbar \tilde{A} \\
 \tilde{A}' &= -\frac{\Gamma}{s} \Abar Q ,
 \end{aligned}
 \label{simpleAQ-linear}
 \end{equation}
 where $\tilde{A} = A-\Abar$.
The solution to system (\ref{simpleAQ-linear}) has wavelength
 \begin{equation}
 X = 2\pi \sqrt{ \frac{6  s^2}{\Gamma \Hbar \Abar f(s)}} =2\pi \sqrt{ \frac{6  s^2}{\Gamma F f(s)} },
 \label{linear_exp}
 \end{equation}
depending on propagation speed $s$ only, in contrast to the wavelength for nonlinear solutions (\ref{period_exp}), which depends also on wave amplitude $\mathcal{A}$.

Figure~\ref{disp_amp_fig} shows the dimensional phase speed $s$ 
and oscillation frequency $\omega(k)=s/k$ of linear and nonlinear waves 
for moist Rossby and MJO modes,
where the wavenumber $k$ is the number of 
waves along the 40,000~km long equator.
With a smaller amplitude, i.e., $\mathcal{A}=0.1$, 
the nonlinear phase speed and oscillation frequency is almost identical to the linear case.
With a larger amplitude, i.e., $\mathcal{A}=0.5$, due to nonlinearity, 
the phase speed drops and so does the frequency.
This analytical result is consistent with two earlier numerical findings.
First, a nonlinear numerical simulation yielded a wave with propagation
speed of 6~m/s, in contrast to the linear wave speed of roughly $7$~m/s \cite{ms11}.
Second, in the stochastic MJO skeleton model, 
the maximal spectral power lies at lower frequencies than the linear
wave frequency \cite{tms14};
since this holds true even when the stochasticity of the model is reduced
(through their parameter $\Delta a$),
it suggests that the reduced frequency is likely due to nonlinear effects 
rather than stochastic effects. 

\subsection{Total energy of traveling waves}
Like the wavelength expression in (\ref{period_exp}), 
the total energy of traveling waves can also be written
as a function of $s$ and $\mathcal{A}$.
For comparing nonlinear waves of different types
(e.g., MJO vs. moist Rossby),
the total energy $\mathcal{E}$ will be the natural quantity to use.

To derive the total energy of the traveling wave,
 it is convenient to first introduce some notations. 
Denote the length of the total domain, 
here, the nondimensional equatoriral circumference, as $L$.
The wavenumber $k$ is then $k = \frac{L}{X}$.  
The conserved energy over the whole domain $L$ is the spatial integral of energy density (\ref{consE_total}):
\begin{equation}
\mathcal{E} = k \int_0^X\ \left[\frac{\tildeq}{2(1-\tildeq)}\left(\frac{Q}{\tildeq}-K-R\right)^2+K^2 + \frac{3}{2}R^2 + \frac{\Hbar}{\Gamma \tildeq} A - \frac{F}{\Gamma \tildeq} \log A\right] \mathrm{d}x
\end{equation}
With (\ref{kr_sol}), the expression is in terms of $Q$ and $A$ only:
\begin{equation}
\mathcal{E} = k \int_0^X\ \left[g(s)Q^2  + \frac{\Hbar}{\Gamma \tildeq} A - \frac{F}{\Gamma \tildeq} \log A \right]  \mathrm{d}x
\end{equation}
where 
\begin{equation}
g(s)=\frac{\tildeq}{2(1-\tildeq)}\left(\frac{1}{\tildeq}-\frac{3s}{(s-1)f(s)} - \frac{6s}{(3s+1)f(s)}\right)^2+\frac{9s^2}{(s-1)^2f^2(s)} + \frac{54s^2}{(3s+1)^2f^2(s)} 
\end{equation}
Next, $Q$ can be eliminated through the Hamiltonian function (\ref{conserveE}):
\begin{equation}
Q^2 = \frac{f(s)}{3\Gamma} \left[\Hbar(\Amax-A) -F(\log \Amax - \log A)\right]
\end{equation}
Hence
\begin{equation}
\mathcal{E} =k \int_0^X \left[\frac{f(s)\ g(s)}{3\Gamma}  \left(\Hbar(\Amax-A) -F(\log \Amax - \log A)\right) + \frac{\Hbar}{\Gamma \tildeq} A - \frac{F}{\Gamma \tildeq} \log A \right]  \mathrm{d}x
\end{equation}
Finally, with the implicit solution (\ref{imp_sol}), 
the integration over $A$ replaces the integration over $x$:
\begin{equation}
\frac{\mathrm{d}x}{\mathrm{d}A} 
= \pm \sqrt{\frac{3s^2}{\Gamma f(s)} } \ A^{-1} \left[\Hbar(\Amax- A)-F(\log \Amax- \log A )\right]^{-1/2}
\end{equation}
This allows us to write 
\begin{equation}
\begin{aligned}
&\mathcal{E} =
2k  g(s) \sqrt{\frac{s^2f(s)}{3\Gamma^3}}\int_{\Amin}^{\Amax}A^{-1}\left[\Hbar(\Amax- A)-F(\log \Amax- \log A )\right]^{1/2} \mathrm{d}A   \\
& + \frac{2k}{\tildeq} \sqrt{\frac{3s^2}{f(s)\Gamma^3}}\int_{\Amin}^{\Amax} A^{-1}\left[\Hbar(\Amax- A)-F(\log \Amax- \log A )\right]^{-1/2}\left(\Hbar A -F \log A \right) \mathrm{d}A  
\end{aligned}
\label{EEs}
\end{equation}

 %
 %
 
%
\begin{figure}
\centering
\includegraphics[width=.8\textwidth]{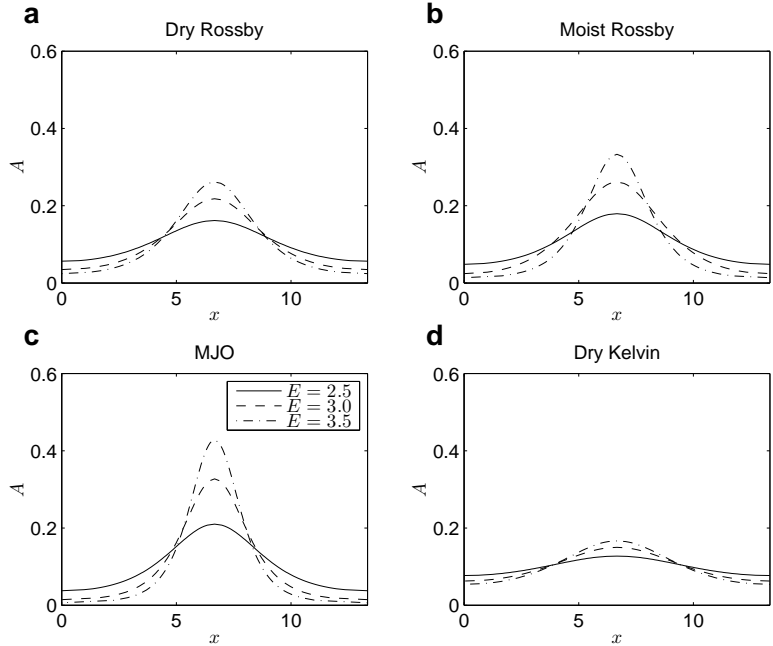}
\caption{Figure of $A$ with different total energy $\mathcal{E}$. Wavenumber $k=2$.}
\label{sol_plot_A}
\end{figure}
\begin{figure}[h]
\centering
\includegraphics[width=.8\textwidth]{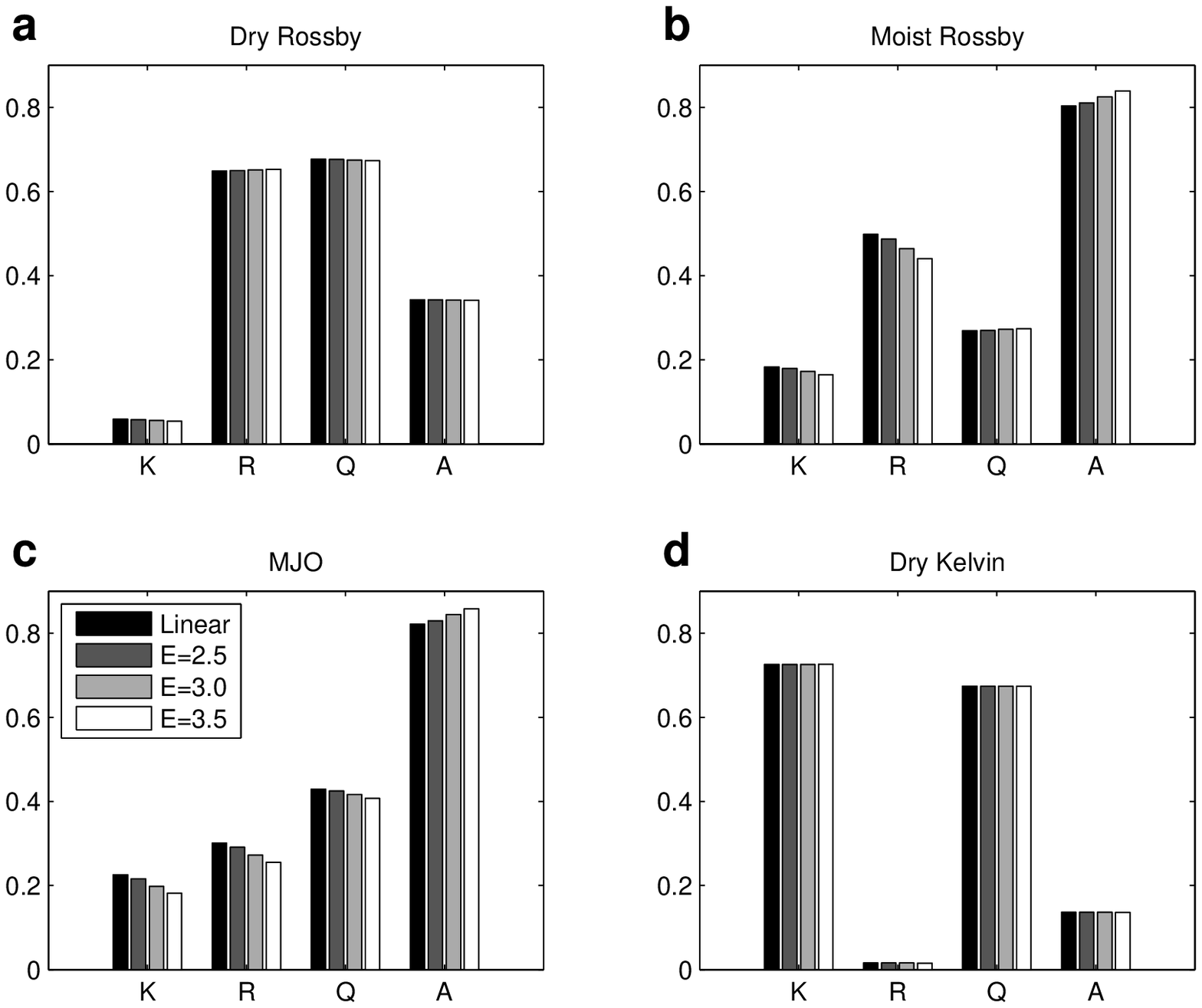}
\caption{Contributions of each component $A$, $Q$, $K$ and $R$ to the linear and nonlinear traveling wave solutions of four modes: dry Rossby, moist Rossby, MJO and dry Kelvin modes. Results for linear waves and nonlinear waves with different total energy $\mathcal{E}=2.5$, $3.0$ and $3.5$ are shown.  Wave number $k=2$. 
Corresponding wave speeds are shown in table~\ref{tb_spd}.}
\label{sol_bar_diffE}
\end{figure}

From (\ref{period_exp}) and (\ref{EEs}), any two from the following four quantities can determine the remaining two: traveling speed $s$, wave amplitude $\mathcal{A}$, wavelength $X$ and total energy $\mathcal{E}$.
For practical reasons, the wavelength $X$ is usually chosen first 
so that the circumference of the equator is a proper domain length.

\begin{table}[h]
\centering
\begin{tabular}{c cccc}
\hline
 & linear & $E=2.5$ & $E= 3.0$ & $E=3.5$\\
\hline
Dry Rossby & -20.96 & -20.87 & -20.70 & -20.56 \\
Moist Rossby & -3.55 & -3.51 & -3.44 & -3.36 \\
MJO & 5.55 & 5.39 & 5.09 & 4.81 \\
Dry Kelvin & 52.30 & 52.29 & 52.26 & 52.24\\
\hline
\end{tabular}
\caption{Traveling wave speeds (in m\ s$^{-1}$) for linear waves and nonlinear waves with different total energy. 
Wave number $k=2$.
Corresponding wave structures are shown in figure~\ref{sol_bar_diffE}.}
\label{tb_spd}
\end{table}

\subsection{Illustrations of waveforms with different amplitudes and energies}
With different wave amplitudes and energies, 
the waveforms have different shapes due to nonlinearity.
To further illustrate the difference between large and small amplitude solutions, 
waveforms with different energies are plotted in figure~\ref{sol_plot_A},
which shows that, as the energy gets larger (so does the amplitude),
the waveforms looks more like a pulse than a sinusoid. 

While the shape of waveforms behaves as a nonlinear feature, 
the relative ratio of variables' amplitudes does not change much with respect to energy.
The ratio is defined by
\begin{equation}
\mathcal{K}: \mathcal{R} : \mathcal{Q}: \mathcal{A},
\end{equation} 
 where $\mathcal{K} = K_{\text{max}}-K_{\text{min}}$ and so on. 
 This ratio is analogous to eigenvectors for the linearized system.
In figure~\ref{sol_bar_diffE}, the ratios are plotted 
with normalizations by L2-norms of vectors
for both nonlinear waveforms with different energies, 
and linear waveforms.
While the variations for moist Rossby and MJO modes are somewhat visible, 
the tendencies for dry modes are hard to see.
Another difference between moist and dry modes that can be seen from
 figure~\ref{sol_bar_diffE} is that,  
 moist modes have the dominant contributions
from the convective envelope $A$,
yet for the dry modes, the greatest contributions are from $R$ or $K$,
depending on whether the mode is dry Rossby or dry Kelvin.
The corresponding wave speeds are given in table~\ref{tb_spd}.

\section{Physical structure} \label{phys-sec}
\begin{figure}
\centering
 \includegraphics[width=.8\textwidth]{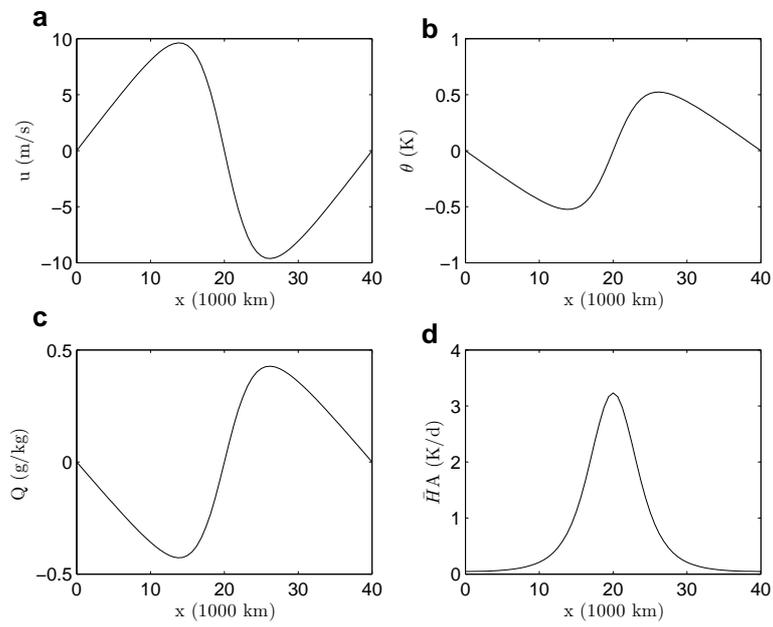}
\caption{Physical quantities at equator ($y=0$) recovered from MJO mode traveling wave solution: zonal velocity $u$, potential temperature $\theta$, moisture $q$ and convective activity envelope $\Hbar A$. Wave number $k=1$. Total energy $\mathcal{E}=3.8$.}
\label{E3_8MJO_figd}
\end{figure}
In this section, several results are presented in
physically relevant terms by returning from the
$(K,R)$ variables to the $(u,v,\theta)$ variables,
by returning from nondimensional to dimensional units,
and by plotting both the zonal ($x$) and meridional ($y$) variations.

First, let's consider the propagation speed $s$ in dimensional units and on a finite domain.
In section~\ref{exist-sec}, the allowed propagation speeds $s$ has four branches: 
dry Rossby, moist Rossby, MJO and dry Kelvin waves with the values marked in figure~\ref{fs-fig}.
From the figure, 
the traveling wave speed~$s$ has a quite large range stretching to infinity for the two dry modes;
in reality, however, the maximum wave speed is confined by two facts which can be illustrated from 
figure~\ref{disp_amp_fig}.

Figure~\ref{disp_amp_fig} shows that for the moisture modes,
the propagation speed $|s|$ decreases as the amplitudes increases, and,
 as the wavenumber increases.
 The same results hold for the dry modes, although not shown here.
 Based on these two facts and based on the fact that 
 wavelengths must be smaller than the Earth's circumference of $\approx$40,000~km,
the maximum traveling wave speeds
are confined by the speed of the $k=1$ equatorial linear waves, for four modes in their absolute value:
\begin{equation}
s_1<s<-\frac{1}{3},\quad s_2<s<0,\quad 0<s<s_3,\quad\text{and}\quad 1<s<s_4.
\label{s-real}
\end{equation}
Here $s_j$ ($j=1,2,3,4$) are  $k=1$ linear traveling wave speeds for four modes:
\begin{equation}
s_1 = -0.62, \quad s_2 =-0.09, \quad s_3 =0.20,  \quad s_4=1.18.
\end{equation}
According to (\ref{s-real}), the dimensional traveling wave speeds are $s\approx$17-31m/s west-propagating for dry Rossby waves,
$s\lesssim$4.7m/s west-propagating for moist Rossby waves, $s\lesssim$10m/s east-propagating for MJO, and $s\approx$50-59m/s east-propagating for dry Kelvin waves.  

Besides propagation speed $s$, other variables are reconstructed. 
With (\ref{reform}), the solution in figure~\ref{E3_8MJO_fig}
converts to physical quantities at equator ($y=0$)  \cite{bm06dao, m03,ms09pnas,ms11, tms14}
 as in figure~\ref{E3_8MJO_figd}.
The wave travels in speed $s\approx 0.18$, or $9.0$~m/s east-propagating.
As shown in figure~\ref{E3_8MJO_figd}, the accumulating moisture leads to active convection after which the moisture drops.
The enhanced convection also occurs at the same phase as the zonal wind converges. 

\begin{figure}
\centering
 \includegraphics[width=.8\textwidth]{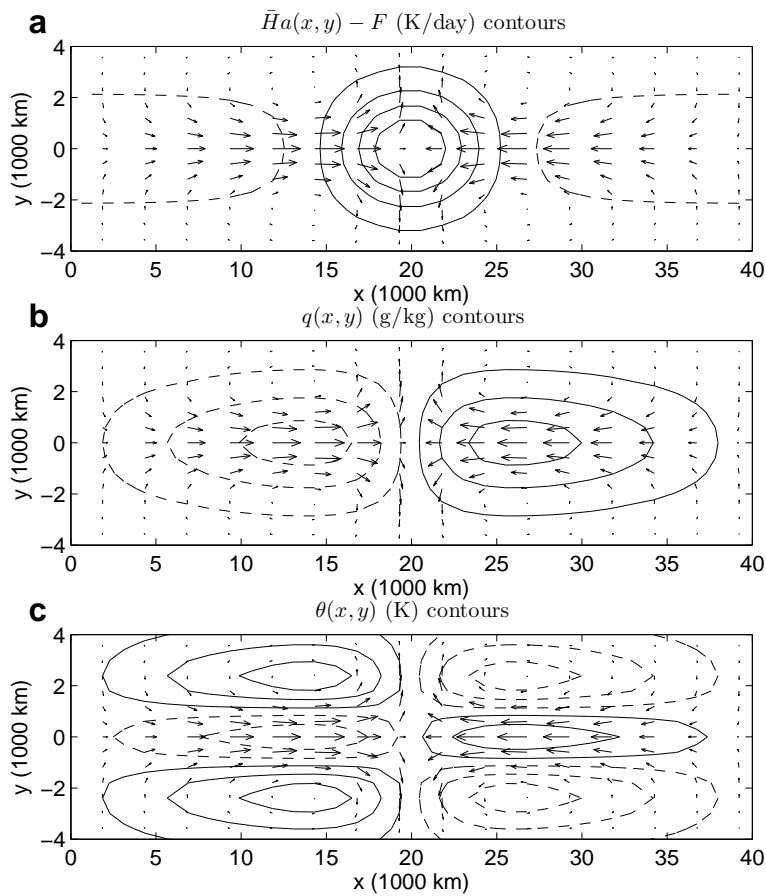}
 \caption{MJO mode traveling wave solution. Total energy $\mathcal{E}=3.8$ and wavenumber $k=1$.
 (a): zonal-meridional structure. Low-level zonal and meridional velocity are shown 
 with contours of the amplitude of the convective activity envelope.
 (b): same as (a), except contours of lower tropospheric moisture, $q(x,y)$.
 (c): same as (a), except contours of lower tropospheric potential temperature anomaly, $\theta$. 
 All positive (negative) contours are shown by solid (dashed) lines.
 For convective heating, moisture, and convergence, 
 the contour intervals are 0.55~K/day, 0.15~g/kg, and 0.24~K, respectively.  
 Maximum zonal and meridional velocities are 9.76~m/s and 0.86~m/s, respectively.}
 \label{mjo_2d_fig}
 \end{figure}

With the reconstructed physical variables at equator, the zonal-meridional structures are recovered 
 in figure~\ref{mjo_2d_fig}
by using the parabolic cylinder functions.
One strongly convective event is present in this case, 
collocated with upward vertical motion and horizontal convergence of the zonal wind.
 Straddling the equator, 
 a pair of anticyclones leads and a pair of cyclones trails the convective activity.
 Also, the maximum lower-troposphere moisture leads the convective maximum.
 Hence, the nonlinear model reproduces the fundamental features of the MJO skeleton model.
 
In figure~\ref{sol_plot_A}, the pulse-like shape of convection envelope for large amplitude $A$ suggests that for strong MJO events,
the enhanced region is narrower than the suppressed region.
This is perhaps realistic due to the
fact that convective activity and precipitation are positive quantities
and hence have negative anomalies that are bounded.
Nevertheless, we are unaware of any observational analysis
that definitively shows this or even targets this question.

\section{Further explorations}\label{future-sec}
This section includes preliminary results for several interesting topics related to the traveling wave solution for the MJO skeleton model.
While the results are not exhaustive, possible directions for future investigations are discussed.
 
\begin{figure}[h!]
\includegraphics[width=.23\textwidth]{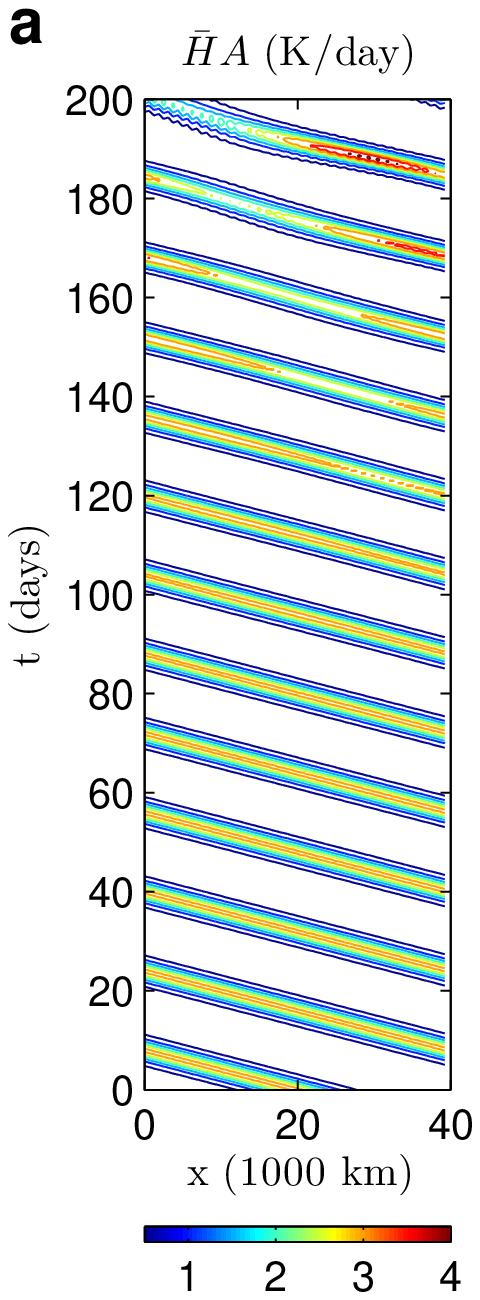}
\includegraphics[width=.23\textwidth]{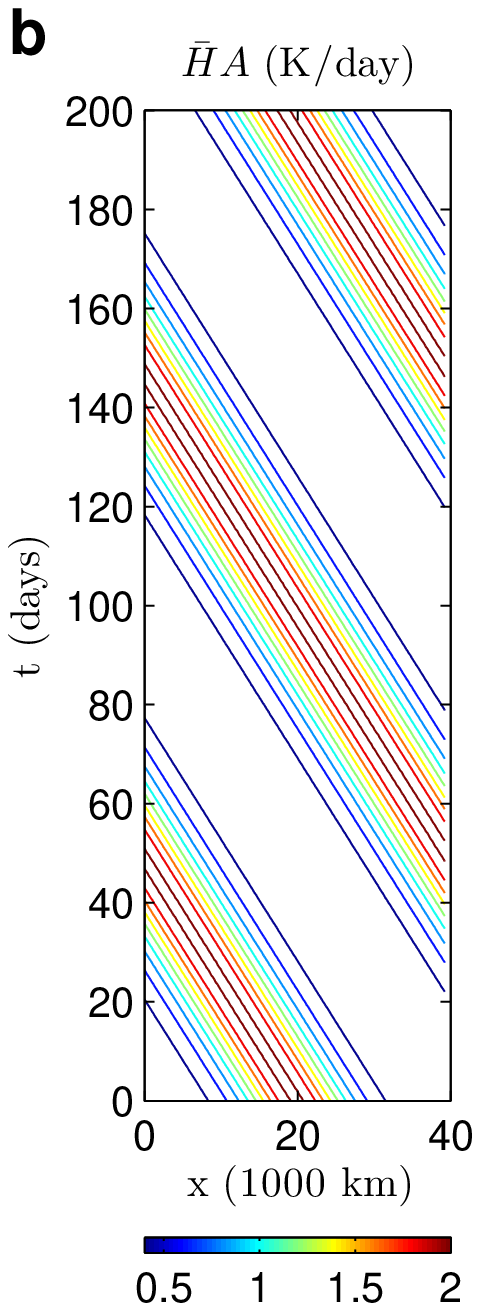}
\includegraphics[width=.23\textwidth]{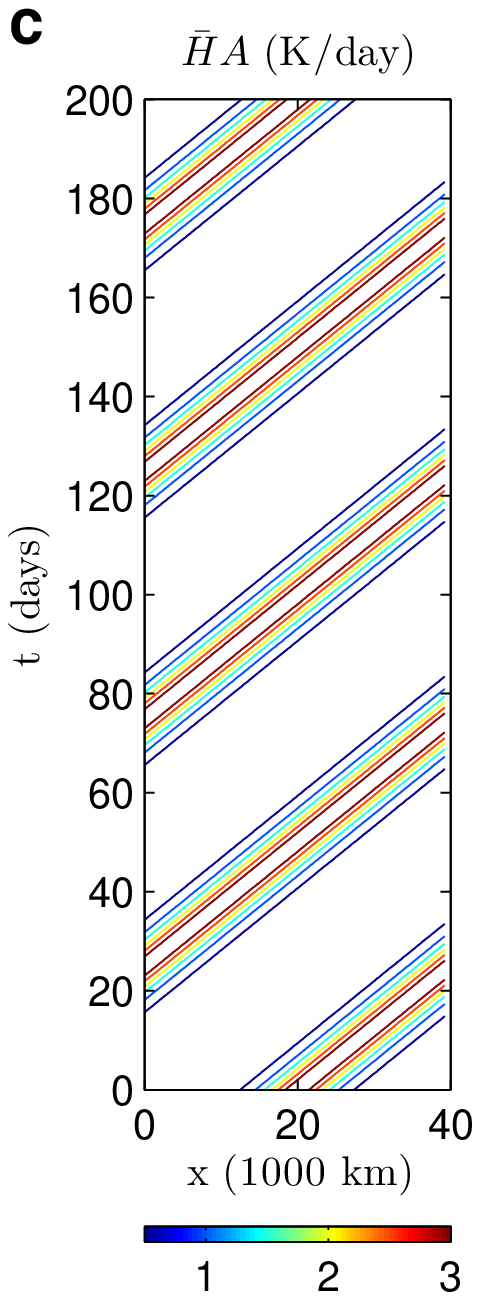}
\includegraphics[width=.23\textwidth]{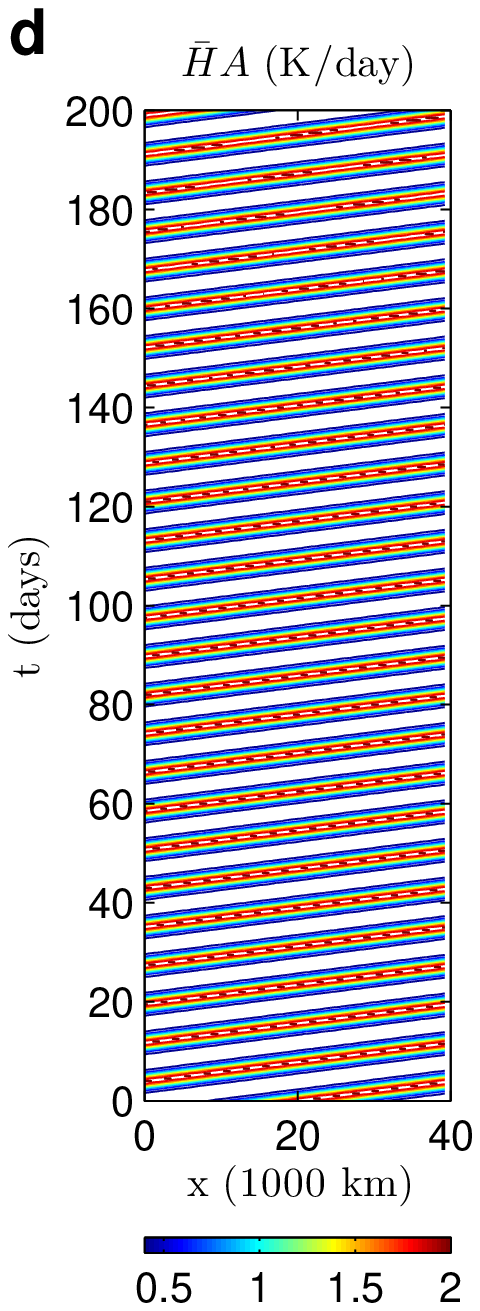}
\caption{Contours of the amplitude of the convective activity envelope, $\Hbar A(x,t)$ at $y=0$. 
(a)-(d) are exact traveling wave solutions for dry Rossby, moist Rossby, MJO and dry Kelvin modes with total energy $\mathcal{E}=3.8$.}
\label{stable_fig}
\end{figure}
\subsection{Stability vs. instability of traveling wave solutions} \label{stab-sec}
Analytical traveling waveforms to system~(\ref{krqa}) are presented in section~\ref{sol-sec}. 
However, it is unclear how the waveforms can be affected by perturbations. 
As a preliminary investigation of the stability/instability of waveforms,
we perform numerical computations for (\ref{krqa}), 
by adopting the scheme from~\cite{ms11}, 
where an operator splitting method is used to separate the linear part (\ref{K_t})-(\ref{Q_t}) and 
nonlinear part (\ref{A_t}).

First, the numerical integrations are initialized with analytical waveforms
 $[K_e,R_e,Q_e,A_e]$
 given by (\ref{imp_sol})-(\ref{kr_sol}).
Tests are run up to $T=200$~days with two energies for four modes: $\mathcal{E}=2.5$ and $\mathcal{E}=3.8$.
Although cases with two energies were performed, 
only results of $\mathcal{E}=3.8$ cases 
are shown as contour plots of convective activity envelope $\Hbar A(x,t)$ in figure~\ref{stable_fig}.
In the figure, the parallel lines are exhibited for most of the plots, 
indicating that
the numerical nonlinear wave is propagating at a fixed speed.

Next, perturbations are placed on the initial waveforms.
The perturbed initial condition is written as
\begin{equation}
[K,R,Q,A]= [K_e,R_e,Q_e,A_e]+[K',R',Q',A'],
\label{pert-desc}
\end{equation} 
where 
the perturbation $[K',R',Q',A']$ are Gaussian functions scaled by $10\%$ of each variable's maximum values,
\begin{equation}
[K',R',Q',A'] = \frac{1}{10} \exp(-\frac{x^2}{16}) [K_{\text{max}}, R_{\text{max}},Q_{\text{max}},A_{\text{max}}],
\label{pert-eq}
 \end{equation}
and they decay fast enough so that the periodicity of the initial condition is not affected.


The numerical integrations are then performed (not shown)
 with perturbed initial data~(\ref{pert-desc}) 
up to $T=200$~days.
To evaluate the effect of perturbations, the quantity ``pattern correlation'' $\mathcal{P}$
is used, which is defined by
\begin{equation}
\mathcal{P}(t)_{[A_e, A] }= \frac{\int A_e(x,t) A(x,t) \mathrm{d}x}{||A_e(x,t)||_2 ||A(x,t)||_2},  
\label{pc_eq}
\end{equation}
where $A_e(x,t)$ and $A(x,t)$ are exact solutions and perturbed solutions.
From (\ref{pc_eq}),
it can be seen that the extreme values of $\mathcal{P}$ are $\pm 1$,
achieved when $A(x,t) = C A_e(x,t)$, for which the sign of $C$ determines
whether it is a maximum or a minimum. 
The proximity of $\mathcal{P}$ to the maximum value $1$
indicates the perturbation has little impact,
thus the solution is stable.

Two cases with total energy $\mathcal{E}=2.5$ and $\mathcal{E}=3.8$
 are performed here,
with initial conditions described in (\ref{pert-desc}).
For traveling waves with total energy $\mathcal{E}=2.5$,
the pattern correlation $\mathcal{P}(t) \ge 97\%$ up to $T=200$~days for all modes.
The number $97\%$ is close to $1$, implying only a slight effect of the perturbation.
In the other case, 
for traveling waves with total energy $\mathcal{E}=3.8$, 
solutions are quite unstable based on the pattern correlation
 except for the moist Rossby mode,
 holding a pattern correlation greater than $97\%$ 
 up to $T=200$~days.
For other three modes, the values of $\mathcal{P}$
 drop significantly within the computational time.
 To identify a great impact from the perturbation, 
 the threshold $90\%$ is used here:
 when the pattern correlation $\mathcal{P}<90\%$,
 the data is greatly affected by the initial perturbation.
 For the other three modes,
dry Rossby mode is the first to have $\mathcal{P}(t)<90\%$, appearing at $T=62$~day;
MJO and dry Kelvin mode have the first $\mathcal{P}(t)<90\%$ around $T=130$~day.
From the numerical experiment performed, 
it may suggest that the moist Rossby mode is most stable,
but no firm conclusion can be drawn based on numerical experiment alone.
%


\subsection{The weak-forcing limit and sech-squared waveforms} \label{asymp-sec}
In section \ref{sol-sec}, the nonlinear traveling wave solution is given implicitly
in terms of an integral for the case of finite forcing, $F$.
In this section, the forcing term, $F$, is taken to be vanishing, i.e., $F\to 0$.
We show now that, in the limiting case, the solution is a solitary wave whose waveform in the function sech$^2$.
According to (\ref{A-Amax}),  
\begin{equation}
\Amax \to \mathcal{A},\quad \Amin \to 0, \quad \text{as }F\to 0.
\end{equation}
Also note that in the integral equation (\ref{ode-now}), the function on the right-hand-side has the asymptotic behavior that 
\begin{equation}
A^2 \left[ \Hbar (\mathcal{A}-A)-F (\log \mathcal{A} - \log A )\right] \sim \Hbar A^2 (\mathcal{A}-A), \quad \text{as } F\to 0. 
\end{equation}
Under this limit, equation (\ref{ode-now}) becomes
 \begin{equation}
 A'=\pm \sqrt{\frac{\Gamma\Hbar f(s)}{3s^2}}A\sqrt{\mathcal{A}-A} 
 \label{ode-solve}
 \end{equation}
 With writing
 \begin{equation}
 h(s) = \sqrt{\frac{\Gamma\Hbar f(s)}{3s^2}},
 \end{equation}
 the solution turns out to be 
  \begin{equation}
A(\tilde{x})=\mathcal{A} \ \text{sech}^2\left[\frac{1}{2}h(s)\sqrt{\mathcal{A}}\ (\tilde{x}-x_0)\right],
\label{sech-sol}
  \end{equation}
  where $x_0$ is an integration constant that determines the location of $\mathcal{A}$.
This integration procedure is similar to get a soliton from KdV equation.

In the solution (\ref{sech-sol}),  the wavelength goes to infinity as the forcing term vanishes, i.e., $F\to 0$.
Besides wavelength, another important length scale is the effective width of $A$, or namely, the length of enhanced convective region, $d$, which is defined as 
\begin{equation}
d = \frac{1}{h(s)\sqrt\mathcal{A}}.
\end{equation}
The effective width $d$ is determined by both the wave amplitude $\mathcal{A}$,
and the wave speed $s$.

\subsection{The model without meridional ($y$) variations}
The model~(\ref{eq1d}) neglects meridional ($y$) variation 
and can be considered as the atmospheric circulation directly above the equator, $y=0$,
 where the Coriolis force is negligible. 
 Similar results hold equally well for (\ref{eq1d}) and (\ref{krqa}).
 While (\ref{eq1d}) neglects important physics, 
 their east-west symmetry simplifies the mathematical formulas.
The model (\ref{eq1d}) would perhaps be easier to study in regard to the further questions 
raised in sections~\ref{stab-sec} and~\ref{asymp-sec}.

%
The key results of traveling wave solutions are provided for system (\ref{eq1d}).
With the traveling wave ansatz $\tilde{x} = x - st$, system (\ref{eq1d}) is 
reduced to the nonlinear oscillatory ODE:
\begin{subequations}
\begin{alignat}{2}
Q' &= \frac{1}{s\left(1-\frac{\tildeq}{1-s^2}\right)}(\Hbar A-F)\label{q-simple}\\
A'&=-\frac{\Gamma}{s}A Q  \label{a-simple}
\end{alignat}
\label{qa-simple-y0}
\end{subequations}
The Hamiltonian function of (\ref{qa-simple-y0}) writes:
\begin{equation}
\mathcal{H} (Q,A)= \frac{1}{s}\left[\frac{\Gamma}{2}Q^2 + \frac{1-s^2}{ 1-s^2-\tildeq }(\Hbar A-F\log{A})\right].
\end{equation}
The solution existence requires that $\mathcal{H}$ has closed contours,
which turn out to be: 
\begin{equation}
\frac{1-s^2}{s^2(1-s^2-\tildeq)}>0,
\end{equation}
or equivalently,
\begin{equation}
s^2>1 \qquad \text{or} \qquad {0<s^2<1-\tildeq}.
\end{equation}
Unlike the system~(\ref{krqa}), the allowed wave propagation speed $s$ for (\ref{eq1d}) has east-west symmetry. 
Here, the limiting values for dry waves and moist waves are
\begin{equation}
 c_{\text{dry}}=1,\quad \text{and } \quad c_{\text{moist}}= \sqrt{1-\tildeq}.
 \label{c_y0}
 \end{equation}
These critical values are the same in the model governing precipitation fronts \cite{fmp04, sm06},
where traveling speed boundaries were set for
 drying fronts, slow moistening fronts and fast moistening fronts.
 Given (\ref{qa-simple-y0})-(\ref{c_y0}), one can derive results
 analogous to those presented in section~\ref{nonlinear-sec}.

\section{Conclusions}
Nonlinear traveling wave solutions were presented for the
MJO skeleton model, and they were compared with their linear counterparts.
The nonlinear traveling waves come in four types,
and the propagation speed of each type is restricted
to lie in a particular interval. 
One wave type has a structure and slow eastward propagation speed
that are consistent with the MJO.
In the nonlinear MJO wave, the convective activity
has a pulse-like shape, with a narrow region of enhanced convection
and a wide region of suppressed convection.  Furthermore,
an amplitude-dependent dispersion relation was derived,
and it shows that the nonlinear MJO has a lower frequency and
slower propagation speed than the linear MJO.  By taking the
small-amplitude limit, an analytic formula was also derived for the
dispersion relation of linear waves.  To derive all of these results,
a key aspect was the model's conservation of energy, which holds
even in the presence of the forcing term, $F$.

The results here suggest several interesting directions for
observational analysis of the MJO.  In particular,
What are the amplitude-dependent properties of the MJO
in observational data?  As one example, for large-amplitude
MJO events, is the region of enhanced convection
stronger and/or narrower than the region of suppressed convection?
An affirmative answer would perhaps be expected due to the
simple fact that convective activity and precipitation are positive quantities
(as illustrated here in figure~\ref{sol_plot_A}).
Nevertheless, we are unaware of any observational analysis
that definitively shows this or even targets this question.
In some numerical simulations of the MJO
\cite{ksmt11,akm13},
it appears to the eye that such an asymmetry might exist 
between enhanced and suppressed convection regions.

Another interesting direction is to investigate the stability
of the nonlinear waves.  In our numerical simulations,
all nonlinear wave types can be reproduced and can propagate
around the Earth's circumference a dozen or more times
with very little change to their structure.
When perturbed, the numerical waves still retain a
significant amount of their coherent propagation,
although some wave modulation can arise, and it can be
difficult to know for certain which perturbed features are part
of the true variability and which are numerical artifacts.
It would be interesting to rigorously prove the stability
or instability of the nonlinear waves.

Finally,
another open question is whether the MJO skeleton model is possibly a
completely integrable system.
The nonlinearity has a mathematical form
that is reminiscent of the Toda lattice model
\cite{t67,t75}
when written in terms of Flaschka's variables
\cite{f74}.
Perhaps one could expand upon this similarity.
The mathematically simplest case to consider is probably
the weak-forcing limit $F\to 0$ on the real line,
rather than the physically realistic case of finite forcing
$F\ne 0$ on a periodic domain.
In the limit of weak forcing,
it was shown that the shape of the nonlinear traveling waves 
is greatly simplified and has a simple sech$^2$ waveform,
the same form as the soliton of the KdV equation.

\medskip
\medskip

{\bf Acknowledgement.}
The authors thank A. Majda for helpful discussions.
The research of S.N.S. is partially supported by the ONR Young Investigator Program through
grant N00014-21-1-0744 and by the ONR-MURI grant N00014-12-1-0912. 
S.C. is supported as a postdoctoral researcher on grant ONR-MURI N00014-12-1-0912.
\medskip

\bibliographystyle{plain}
\bibliography{mjobib.bib}{}

%
%
%
%
%
%
\end{document}